\documentclass[12pt]{article}
\usepackage{pic04}
\usepackage{hyperref}
\usepackage{url}
\usepackage{graphicx}

\begin{document}

\title{\bf FLAVOUR CHANGING NEUTRAL CURRENT B DECAYS AT BABAR}
\author{
Fabio Bellini    \\
{\em Universit\`a degli Studi di Roma ``La Sapienza'', Piazzale A.~Moro 5 Rome, 00185}}
\maketitle

\baselineskip=14.5pt
\begin{abstract}
Recent BaBar results on rare B decays involving flavour-changing neutral
 currents are presented.
New measurements of the CP asymmetries in $b \to s \gamma$ decays are reported as well as $b \to s l^+l^-$ branching ratio measurements\footnote{Charge conjugate modes will be assumed throughout this text}.
\end{abstract}

\baselineskip=17pt

\section{Introduction}

Since the first measurement of the exclusive $B \to K^* \gamma$ decay rate by
CLEO \cite{bib:cleo-first-btosgamma}, rare $B$ decays involving
 flavour changing neutral current have been a unique probe to search for new physics.
 In the Standard Model (SM), the lowest order diagram for the $b \to s \gamma$ decay is a loop (radiative penguin) diagram of top quark and $W$ boson.  In principle, new
particles such as charged Higgs or SUSY partners can form the same loop
diagram and may modify the SM amplitude.  A comparison between the
measured rate and the SM prediction has provided a stringent constraint
on such new particles.  
The inclusive decay is to date accurately calculated up to the
next-to-leading order QCD corrections \cite{Misiak} and several measurements have 
already been performed.
On the contrary, exclusive measurements, do
not give a further constraint to new physics because of the large uncertainties in the form factors computation.
Other observables, such as the partial rate asymmetry ($A_{CP}$) between
charge conjugate modes, could be more useful to constrain new physics.  For
example, the SM predicts very small asymmetry ($\approx$1\%)\cite{lunghi}, while there are several
extensions of the SM predicting much larger  $A_{CP}$. 
Electroweak processes, like $b \to s l^+l^-$, are also useful probes
for new physics searches.  Expected rates are two order of magnitude
smaller than for $b \to s \gamma$. At the lowest order, the decay is described
by an electroweak ($Z$) penguin diagram and a $W$-box diagram in
addition to the radiative penguin.  If new particles with large weak bosons couplings exist, one can expect some
additional contributions to $b \to s l^+ l^-$ that are not visible in
$b \to s \gamma$.

\section{Measurements of CP Asymmetries in $b\to s \gamma$}

The exclusive $B\to K^*\gamma$ analysis is detailed in \cite{Aubert:2004te}.
The $K^*$ is reconstructed in the four modes $K^{*0}\to K^+ \pi^-,K^0_S \pi^0$ and $K^{*+}\to K^+ \pi^0,K^0_S \pi^+$ and combined with a high energy isolated photon to form a B meson candidate. The background, mostly from continuum production, is suppressed by means of event topology variables.
To reject events in which the photon comes from a $\pi^0(\eta)$ decay, a veto on $m_{\gamma\gamma}$ invariant mass is applied.
The discriminating variables are the beam energy-substituted 
mass 
$m_{ES}=\sqrt{(E_{beam}^*)^2 -(\vec p_B^*)^2}$  and the energy difference 
$\Delta E^*=E^*_B -E^*_{beam}$, where  $\vec p_B$, $ E^*_B$ and $E^*_{beam}$ 
denote the B-momentum, B-energy and beam energy in the center-of-mass 
(CM) frame, respectively.
The signal yields are extracted from a likelihood fit to $m_{ES}$ and $\Delta E$.
A ``semi-inclusive'' method, which measures a sum of exclusive $B \to X_s\gamma$ decays, is also used in BaBar~\cite{Aubert:2004hq}.
The hadronic system $X_s$ is reconstructed in 12 final states 
including a $K^0_S$ or a $K^+$ and up to three pions (at most one $\pi^0$).
The signal yields are extracted from a likelihood fit to $m_{ES}$. 
The results for the direct CP asymmetry, based on $81.9 fb^{-1}$, are reported in Table \ref{tab1}. These are consistent with zero and statistics limited. 

\begin{table}
\centering
\caption{ \it $A_{CP}$ measurements in $b\to s\gamma$. Errors are statistics and systematics respectively.}
\vskip 0.1 in
\begin{tabular}{|l|c|c|c|} \hline
       & $B \to K^* \gamma$ & $B \to X_s\gamma$ \\
\hline
\hline
Direct CP asymmetry $A_{CP}$ & (-1.3 $\pm$3 $\pm$1)\% & (2.5 $\pm$5.0 $\pm$1.5)\% \\ 
\hline
\end{tabular}
\label{tab1}
\end{table}

\section{Measurements of Branching Fractions in $b\to s l^+ l^-$}


The exclusive $b\to s l^+ l^-$ measurement is described in \cite{bp35}.
Eight final states are reconstructed where a $
K^+, K^0_S, K^{*0}$ or $K^{*+}$ recoils
against a $\mu^+ \mu^-$ or $e^+ e^-$ pair, using an integrated luminosity of 
$ 113.1\ \rm fb^{-1}$. Specific selection criteria are used to suppress individual backgrounds. 
Event shape variables are used to eliminate the continuum background.
To reject events from $B \rightarrow J/\psi (\psi(2S)) K^{(*)}$ decays with 
$J/\psi (\psi(2S)) \rightarrow l^+ l^-$, a veto on $m_{ll}$ is applied. 
In each of the four $K^{(*)} l^+ l^-$ final states,
a signal is extracted from a fit to the $m_{ES}, \Delta E, (m_{k \pi})$ distributions.
\begin{table}
\centering
\caption{ \it Branching Fractions (BF) Predictions and Measurements in $b\to s l^+ l^-$ decays. Errors are statistics and systematics respectively.}
\begin{tabular}{|l|c|c|c|} \hline
  & $B\to K l^+ l^-$ & $B \to K^* l^+ l^-$  & $B \to X_s l^+ l^-$\\
\hline
\hline
BF Predictions ($\times 10^7$)& $3.5\pm 1.2$& $11.9\pm 3.9$ & $42\pm 7$ \\
Measured BF ($\times 10^7$) & $6.5^{+1.4}_{-1.3}\pm 0.4$ & $8.8^{+3.3}_{-2.9}\pm 1.0$ & $ 56 \pm15 \pm12$ \\ 
\hline
\end{tabular}
\label{tab2}
\end{table}
In the ``semi-inclusive'' approach \cite{Aubert:2004it}, the reconstructed hadronic system $X_s$ consists of one $K^0_S$ or  $ K^+$ and up to two pions (at most one $\pi^0$). 
Background rejection is similar to the exclusive analysis. The signal yield is extracted from a likelihood fit to $m_{ES}$ and the analysis is performed with an integrated luminosity of  $ 81.9\ \rm fb^{-1}$. The measured branching ratios are reported in Table \ref{tab2} togheter with the theoretical predictions~\cite{Alisll}.
These results are in agreement with the SM predictions within the current level of accuracy and have errors comparable to the theoretical prediction precision. With larger data samples at the {\it B-factories}, it will be possible to make precise tests of the theoretical predictions for the differential distributions in these decays.

\end{document}